\documentclass[aps,prd,nofootinbib,superscriptaddress]{revtex4}
\usepackage[sumlimits]{amsmath} 
\usepackage{epsfig}
\usepackage{amssymb}
\usepackage{graphicx}
\usepackage{pstricks}
\usepackage{color}
\usepackage{bbold}
\usepackage{hyperref}
\usepackage{yfonts}

\begin{document}

%
%
\title{Geometrical scalar back-reaction effects in inflation}

\author{Rafael Hern\'{a}ndez-Jim\'{e}nez}
\email{rafaelhernandezjmz@gmail.com}
\author{Claudia Moreno} 
\email{claudia.moreno@academico.udg.mx}
\affiliation{Departamento de F\'isica,
CUCEI, Universidad de Guadalajara\\
Blvd. Marcelino García Barragán 1421, C.P. 44430, Guadalajara, Jalisco, M\'exico}
%
%
\date{\today}

%
%
\bigskip
\begin{abstract}
Starting with the Lagrangian formulation of General Relativity, we will conduct an investigation into the production of spacetime waves, due to a geometric boundary term of a closed extended manifold, within the tensor and scalar sectors. This scheme will be studied in an inflationary universe. We explore two distinct scenarios: Cold Inflation and Warm Inflation. The scalar modes $Z_{k}^{\mathbb{R}}$ and $Z_{k}^{\mathbb{I}}$ oscillate within the horizon, and they become constant at (or right after) horizon crossing $k\simeq aH$ and they remain so when radiation starts to dominate. The larger $k/k_{0}$ the $Z_{k}$'s amplitudes increase too. In general we can notice that radiation reduces the size of the $Z_{k}$'s amplitudes, hence yielding smaller signals of such modes. The tensor sector shows an irregular journey due to their abruptly growth just as they cross the horizon. This upshot hinders any probable observational hint or signal. However, we expect this novel mechanism of spacetime waves production brings new cosmological sources, for which no astrophysical source has been identified. 
\end{abstract}


\maketitle

%
%
\section{Introduction}
Early universe cosmology is certainly full of stimulating and exotic physics. Indeed, we need enhanced tools to describe such framework. To begin with the Lagrangian formulation of General Relativity (GR) from the Einstein-Hilbert (EH) action, formed with the only independent scalar constructed from the metric, which is no higher than second order in its derivatives, the Ricci scalar $R = g^{\mu\nu}R_{\mu\nu}$ times the square root of negative determinant of the metric tensor $\sqrt{-g}$. Then, the equations of motion should come from varying the action with respect to the metric $g^{\mu\nu}$. Hence, the variation of the Ricci tensor $\delta R_{\mu\nu}$ yields the covariant divergence of a vector, which by Stokes's theorem, this is equal to a boundary contribution at infinity which we can set to zero by making the variation vanish at infinity. However, when the underlying spacetime manifold has a boundary $\partial V$, aforementioned procedure leads to a troublesome assumption. To solve this inelegant issue Hawking-Gibbons-York (HGY) proposed to add a counter term in the EH action, which relates the boundary constraint and extrinsic curvature \cite{Gibbons:1976ue, York:1972sj}, to in fact cancel such input. Nonetheless, by eliminating this extremum, any physical phenomena at the border are excluded, such as back-reaction effects, provided these are indeed considered irrelevant. Instead of dropping the boundary expression, an alternative proposal is to take them into account as a physical source of geometric nature \cite{Ridao:2015oba, Ridao:2014kaa}. The contribution of these can be associated to spacetime waves (SWs) production. We will illustrate this new scheme for an inflationary universe. A clever mechanism that generates the primordial seeds required to explain the observed large scale structure, apart from solving the horizon and flatness problems \cite{Starobinsky:1980te, Sato:1980yn, Guth:1980zm, Albrecht:1982wi, Linde:1981mu}. 

The inflationary paradigm is mainly characterised by the existence of an homogeneous scalar field, the inflaton $\phi$, which is the cause of an abrupt exponential expansion; and the presumed presence of quantum fluctuations $\delta\phi$ around $\phi$. In the standard inflationary picture, called cold inflation (CI), the super fast period of accelerated expansion quickly dilutes away all traces of any pre-inflationary matter or radiation density, leaving this framework in a vacuum state. However, this yields a supercooled universe. So in order to successfully explain the transition from inflation to the Standard Big Bang (SBB) scenario, and hence the physics of recombination leading to the Cosmic Microwave Background (CMB) that we observe today, the inflaton energy density necessarily requires to be transferred into ordinary matter and radiation, and thus to its interactions with other fields. Accordingly, the inflaton field could be coupled to other components and might dissipate its vacuum energy and warm up the universe. This alternative scenario is known as the warm inflation (WI) paradigm \cite{Berera:1995wh,Berera:1995ie,Berera:1996fm,Berera:2008ar}, where dissipative effects and associated particle production can, in fact, sustain a thermal bath concurrently with the accelerated expansion of the universe during inflation. Hence, this augmented mechanism becomes phenomenological more relevant. Certainly, WI has attractive characteristics. For instance, even if radiation is subdominant during inflation, it may smoothly become the leading component at the end of inflation, with no need for a separate reheating or preheating period. Moreover, dissipation also affects scalar perturbations, albeit for WI the fluctuations of the inflaton are thermally induced, which are already classical upon definition, hence they may bring an interpretation of the nature of the classical inhomogeneities observed in the CMB; consequently, there is no need to explain the troublesome quantum-to-classical transition problem of CI, due to the purely quantum origin of the CI density perturbations. Importantly, it has been observed that many scenarios of WI are within the observable window \cite{Bastero-Gil:2016qru, Bastero-Gil:2018uep, Bastero-Gil:2018yen}. Moreover, it is been shown that WI can dodge the proposed swampland conjectures \cite{Motaharfar:2018zyb, Das:2018rpg, Bastero-Gil:2018yen}. Thus, WI is in accordance with both the current cosmological observations and the proposed Swampland Criteria. 

In the context of CI it is expected the generation of Gravitational Waves (GWs)~\cite{Starobinsky:1979ty, Rubakov:1982df, Fabbri:1983us, Abbott:1984fp}. Indeed, they are proposed as the mechanism that could polarised the CMB, inducing a specific gradient-free “B-mode” pattern \cite{Kamionkowski:1996zd, Seljak:1996gy, Kamionkowski:1996ks, Zaldarriaga:1996xe, Seljak:1996ti}. Other studies have been done introducing more components to inflation to analyse GWs creation; for instance, primordial magnetic fields \cite{Sharma:2019jtb}. On the other hand, the research of this phenomena within a WI scenario has been recently explored in~\cite{Arya:2019wck,Bastero-Gil:2021fac}. For instance, Bastero-Gil and Díaz-Blanco~\cite{Bastero-Gil:2021fac} found an enhancement of GWs production caused by the presence of light scalars couple to the inflaton. Apart from this standard framework, GWs production due to a closed Riemann manifold have been in fact studied before by \cite{Ridao:2015oba, Ridao:2014kaa}. And recently, authors in \cite{Bellini:2022mlh,Bellini:2023gct,Bellini:2023zev} have studied Spacetime Waves (SWs) generation in preinflation with extended GR. Although in both contexts they calculated the SWs dynamics of a primordial CI stage. However, we envisage a richer phenomenological description of such early universe physics. We will conduct an investigation into the production of these gravitational ripples due to the existence of a closed manifold, and its geometric boundary terms, within two frameworks: CI and WI.  

This paper is organized as follows. In section~\ref{back-reaction_dynamics} we briefly present the back-reaction dynamics on an extended manifold within a homogeneous and isotropic flat Friedmann-Lema\^{i}tre-Robertson-Walker Riemann background. In section~\ref{SW-CI}, we study SWs production in a CI scenario. In section~\ref{SW-WI}, we analyse SWs within a WI paradigm, we highlight how this enhanced mechanism in fact changes the evolution of our relevant parameters. Finally, in section~\ref{conclusions} we will give the conclusion and outlook of this work.

%
%

\section{Back-reaction dynamics on an extended manifold}\label{back-reaction_dynamics}
Einstein field equations can be obtained varying the EH action, where the usual setting does not consider the boundary terms. This scheme does describe closed physical systems that do not interact with the extrema, hence no back-reaction effects, due to them, are taken into account. However, to include them we must consider the border contribution of a closed manifold. In this work we will describe the consequences of such effects but within an extended GR, that is the background system is represented by a Riemann manifold, and the geometric fluctuations (due to the boundary) are placed into an extended manifold. We begin with the EH action described gravitation and matter in our universe and it is represented by:
\begin{equation}\label{general_action}
\mathcal{S}=\int d^{4}x\sqrt{-g}\left[\frac{R}{2\kappa}+\mathcal{L}_{m} \right]\,,
\end{equation}
where $\kappa = 8\pi G$~{\footnote{In the entire paper we use natural units: $c=1$.}}, $g$ is the determinant of the covariant background tensor metric $g_{\mu\nu}$, $R=g^{\mu\nu}R_{\mu\nu}$ and $R^{\alpha}_{\,\mu\nu\alpha}=R_{\mu\nu}$ are the scalar curvature and Ricci curvature tensor, respectively. They are derived from the curvature tensor $R^{\alpha}_{\,\beta\gamma\delta}= \Gamma^{\alpha}_{\,\beta\delta\,,\gamma} - \Gamma^{\alpha}_{\,\beta\gamma\,,\delta} + \Gamma^{\epsilon}_{\,\beta\delta}\Gamma^{\alpha}_{\,\epsilon\gamma} - \Gamma^{\epsilon}_{\,\beta\gamma}\Gamma^{\alpha}_{\,\epsilon\delta}$, where the Christoffel symbols are written in terms of the metric tensor and its partial derivatives $\Gamma^{\sigma}_{\alpha\beta} = \left(g_{\gamma\beta\,,\alpha}+g_{\gamma\alpha\,,\beta}-g_{\alpha\beta\,,\gamma}\right)g^{\sigma\gamma}/2$. The Greek indices run from 0 to 3, additionally if latin indices $i$, $j$, etc. appear, they go from 1 to 3. Finally, $\mathcal{L}_{m}$ is an arbitrary Lagrangian density which describes matter. First of all, let us recall that the standard procedure to obtain the Einstein field equations one must vary $\mathcal{S}$ with respect to metric tensor, and in fact let us consider variations with respect to the inverse metric $g^{\mu\nu}$. Accordingly, we have the variation of the action matter
\begin{equation}\label{variation_Lm}
\delta\left[\sqrt{-g}\mathcal{L}_{m}(g^{\mu\nu},g^{\mu\nu}_{\,\,\,,\lambda})\right] = - \frac{\sqrt{-g}}{2}\delta g^{\mu\nu} T_{\mu\nu} \,,  
\end{equation}
here, we have used the generic definition of the stress-energy tensor:
\begin{equation}\label{stress-energy_tensor1}
T_{\mu\nu} = g_{\mu\nu}\mathcal{L}_{m} - 2\frac{\delta\mathcal{L}_{m}}{\delta g^{\mu\nu}}  \,.  
\end{equation}
Now, we consider the gravitational action where its variation is
\begin{equation}\label{vary_lagrange}
\delta\left[\sqrt{-g}R\right] = \sqrt{-g}\left[\delta g^{\alpha\beta}\, G_{\alpha\beta}+g^{\alpha\beta}\delta R_{\alpha\beta}\right] \,,   
\end{equation}
where $G_{\alpha\beta}=R_{\alpha\beta} - g_{\alpha\beta}R/2$ is the Einstein tensor. By considering the fluctuations (as a geometric response to some physical field fluctuations) $\delta g_{\alpha\beta}$ as the origin of curvature fluctuations ~\cite{Ridao:2014kaa, Ridao:2015oba, Hernandez-Jimenez:2022daw}, i.e.
\begin{equation}\label{variation_Ricci}
\delta R_{\alpha\beta} = \Lambda(x)\,\delta g_{\alpha\beta} \,. 
\end{equation}
Thus, the dynamics of the system with boundary conditions included will be given by the Einstein equations (provided that $\delta\mathcal{S}=0$), that now takes the form:
\begin{equation}\label{Einstein-eqs}
G_{\alpha\beta} + \Lambda \, g_{\alpha\beta} =  \kappa T_{\alpha\beta} \,.
\end{equation}
Accordingly, {\it{the flow of the fluctuations of some physical field would be the origin of the cosmological constant on large (cosmological) scales}.} In other words, $\Lambda(x)$ appears as a response to the inclusion of a finite boundary in the Lagrangian formulation of GR~\cite{Ridao:2014kaa, Ridao:2015oba, Hernandez-Jimenez:2022daw}. On the other hand, we will present definitions and deductions of new covariant derivatives on an extended manifold. We follow the prescription presented in~\cite{Bellini:2022mlh,Bellini:2023gct,Bellini:2023zev, Morales:2021hgj}. First, the variation of the Ricci tensor $\delta R_{\alpha\beta}$ on the extended manifold can be defined as an extended version of the Palatini identity \cite{PalatiniDeduzioneID,Guarnizo:2010xr}:
\begin{equation}\label{delta_Ricci}
\delta R^{\gamma}_{\,\,\alpha\beta\gamma} = \delta R_{\alpha\beta} = \frac{M_{P}}{b} \left[\left(\delta\Gamma^{\gamma}_{\,\,\alpha\gamma}\right)_{||\beta} - \left(\delta\Gamma^{\gamma}_{\,\,\alpha\beta}\right)_{||\gamma} \right]  \,,
\end{equation}
where ($||$) denotes the covariant derivative on the extended manifold with self-interactions included. Here $b$ is a dimensionless parameter to be determined, and $M_{P}=1/\sqrt{8\pi G}$ is the reduced Plank mass. To describe the displacement of the extended manifold: $\delta\Gamma^{\gamma}_{\,\,\alpha\beta}$ with respect to the semi-Riemann manifold, we shall consider the Levi-Civita connections:
\begin{equation}
\tilde{\Gamma}^{\gamma}_{\,\,\alpha\beta} = \Gamma^{\gamma}_{\,\,\alpha\beta} + \delta\Gamma^{\gamma}_{\,\,\alpha\beta} = \Gamma^{\gamma}_{\,\,\alpha\beta} + \frac{b}{M_{p}}\sigma^{\gamma}g_{\alpha\beta} \,, 
\label{conection}
\end{equation}
such that $\delta\Gamma^{\gamma}_{\,\,\alpha\beta} = M_{P}^{-1}\, b\,\sigma^{\gamma}g_{\alpha\beta}$ takes into account the perturbed geometry of space-time with respect to the Riemann one. Also, $\sigma$ is a scalar field that describes the scalar back-reaction of geometry due to the perturbations of a given matter sector; we shall denote the partial derivative $\sigma_{,\gamma}\equiv\sigma_{\gamma}=\partial\sigma/\partial x^{\gamma}$ (subsequently $\sigma^{\gamma}=\partial\sigma/\partial x_{\gamma}$). The perturbations will be considered finite, but they can be large, since our formalism is {\it non-perturbative}. The closed 3d-hypersurface is bounded and can be defined on any region of the background manifold. If that background is spatially isotropic and homogeneous, the results obtained on a given region of space-time will be valid everywhere else. Now, we shall consider that SWs are related to $\sigma$ by:
\begin{equation}\label{sigma}
\sigma = g^{\alpha\beta}\delta\Psi_{\alpha\beta} \,,
\end{equation}
where $\delta\Psi_{\alpha\beta}$ is a symmetric 2-rank tensor encoding the sector that describes the SWs. Then the varied connection can be written in terms of the tensor components according to eq.~\eqref{conection} as:
\begin{equation}
\delta\Gamma^{\mu}_{\theta\epsilon} = \frac{b}{M_{P}}\left(g^{\alpha\beta}\nabla^{\mu}\delta\Psi_{\alpha\beta}\right)g_{\theta\epsilon} \,,
\end{equation}
where $\nabla_{\mu}g_{\alpha\beta}=0$ (hence $\nabla^{\mu}g^{\alpha\beta}=0$) is the covariant derivative of the metric tensor on the Riemann manifold. We define the covariant derivative of the metric tensor on the extended manifold, with self-interactions represent by the parameter $\xi$, as:
\begin{equation}
g_{\alpha\beta||\mu} = \nabla_{\mu}g_{\alpha\beta} - \delta\Gamma^{\nu}_{\,\,\alpha\mu}g_{\nu\beta} - \delta\Gamma^{\nu}_{\,\,\beta\mu}g_{\alpha\nu} + 2\xi\sigma_{\mu}g_{\alpha\beta}\,.
\end{equation}
Then, we can define the variation of the metric tensor eq. \eqref{variation_Ricci} on the extended manifold, as:
\begin{equation}
\delta g_{\alpha\beta} = g_{\alpha\beta||\mu} U^{\mu} \,,   
\end{equation}
where $U^{\mu}=dx^{\mu}/ds$ are the components of the
relativistic velocity of an observer that moves on the Riemann manifold. Therefore the variation of the metric tensor can be written in terms of the SWs components, $\delta\Psi_{\gamma\theta}$:
\begin{equation}
\delta g_{\alpha\beta} = g^{\gamma\theta}\left\{2\xi g_{\alpha\beta}U^{\mu}\nabla_{\mu}\delta\Psi_{\gamma\theta} - \frac{b}{M_{P}}\left[U_{\alpha}\nabla_{\beta}\delta\Psi_{\gamma\theta} + U_{\beta}\nabla_{\alpha}\delta\Psi_{\gamma\theta} \right]\right\} \,. 
\end{equation}
We shall consider that the relativistic velocity $U^{\alpha}$ does not change on the extended manifold, i.e. $\delta U^{\alpha}=0$, therefore:
\begin{equation}
\delta U^{\alpha} = U^{\beta}U^{\alpha}_{\,\,||\beta} = \delta U^{\beta}\left(\nabla_{\beta}U^{\alpha} + \delta\Gamma^{\alpha}_{\,\,\beta\nu}U^{\nu} - \xi\sigma_{\beta}U^{\alpha}\right) \,.   
\end{equation}
This means that the evolution of $U^{\alpha}$ will be described by:
\begin{equation}
U^{\beta}\nabla_{\beta}U^{\alpha} = - \delta\Gamma^{\alpha}_{\,\,\beta\nu}U^{\nu}U^{\beta} + \xi\sigma_{\beta}U^{\alpha}U^{\beta} \,,    
\end{equation}
where, by using the fact that $U^{\alpha}U_{\alpha}=1$, we obtain the geodesic equation:
\begin{equation}
\frac{dU^{\alpha}}{ds} + \Gamma^{\alpha}_{\,\,\beta\nu}U^{\beta}U^{\nu} = -\frac{b}{M_{P}}\sigma^{\alpha} + \xi\sigma_{\beta}U^{\alpha}U^{\beta} \,,    
\end{equation}
which in terms of the SWs components becomes:
\begin{equation}\label{geodesic}
\frac{dU^{\alpha}}{ds} + \Gamma^{\alpha}_{\,\,\beta\nu}U^{\beta}U^{\nu} = g^{\gamma\theta}\left\{ \xi U^{\alpha}U^{\beta}\nabla_{\beta}\delta\Psi_{\gamma\theta} - \frac{b}{M_{P}}\nabla^{\alpha}\delta\Psi_{\gamma\theta} \right\} \,.   
\end{equation}
Note that the source term (the right hand of eq.~(\ref{geodesic})) is given by the flux of the tensor sector $\nabla^{\alpha}\delta\Psi_{\gamma\theta}$. To calculate the dynamic equations of $\sigma$ we must use the eqs.~(\ref{variation_Lm}, \ref{vary_lagrange}), from which we obtain:
\begin{equation}
\left[\frac{b}{M_{P}}\left(\sigma^{\beta}U^{\alpha} + \sigma^{\alpha}U^{\beta}\right) - 2\xi g^{\alpha\beta}\sigma_{\mu}U^{\mu}\right]\left[G_{\alpha\beta} - \kappa T_{\alpha\beta}\right] = 3\left[\Box\sigma + \left(\frac{2b}{M_{P}} + \xi\right)\sigma_{\mu}\sigma^{\mu}\right] \,,    
\end{equation}
where $\Box\sigma=g^{\alpha\beta}\nabla_{\alpha}\nabla_{\beta}\sigma$ is the d'Alembertian operator, and we have used the expression
\begin{equation}
\delta g^{\alpha\beta} = g^{\alpha\beta||\mu} U^{\mu} = \frac{b}{M_{P}}\left(\sigma^{\alpha}U^{\beta} + \sigma^{\beta}U^{\alpha}\right) - 2\xi\sigma_{\mu}U^{\mu}g^{\alpha\beta} \,.
\end{equation}
To describe a linear wave equation for $\sigma$, we must fix the gauge $\xi=-2b/M_{P}$; then we consider eqs.~(\ref{variation_Ricci}, \ref{Einstein-eqs}), having as a result:
\begin{equation}\label{equation_sigma}
\Box\sigma = \frac{6b\,\Lambda}{M_{P}}\sigma_{\mu}U^{\mu} \,.    
\end{equation}
Due to the self-interaction of back-reaction effects, the source of $\sigma$ (right hand of eq.~(\ref{equation_sigma})) depends on itself. Furthermore, the source is viewed differently by distinct observers, depending of how they are moving with respect to it, since the relativistic velocity $U^{\mu}$ is presented explicitly. On the other hand, making use of the eqs.~(\ref{sigma}, \ref{equation_sigma}), we obtain an exact equation for the SWs:
\begin{equation}\label{equation_deltaPSI}
\Box\delta\Psi_{\alpha\beta} = \frac{6b\,\Lambda}{M_{P}} U^{\mu}\nabla_{\mu}\delta\Psi_{\alpha\beta} \,,    
\end{equation}
which is a wave differential equation for $\delta\Psi_{\alpha\beta}$. Note that the source term (right hand of eq.~(\ref{equation_deltaPSI})) is originated by the flow through the 3d-closed hypersurface when the action is varied. Moreover, eq.~(\ref{equation_deltaPSI}) is valid on an arbitrary curved background spacetime. Here, it is important to remark that:
\begin{equation}
\Box\delta\Psi_{\alpha\beta} = g^{\alpha\beta}\nabla_{\alpha}\nabla_{\beta} \delta\Psi_{\alpha\beta} \,,   
\end{equation}
where $\nabla_{\alpha}\nabla_{\beta} \delta\Psi_{\alpha\beta}$ must be taken as a covariant derivative $\nabla_{\alpha}$ of a 3-rank tensor: $\nabla_{\beta} \delta\Psi_{\alpha\beta}$. In the next subsection we will present all relevant equations that arise from a spatially isotropic and homogeneous spacetime background. 
\subsection{Spatially isotropic and homogeneous spacetime background}\label{FLRW-model}
We will work within a model described by a perfect fluid with a homogeneous and isotropic flat Friedmann-Lema\^{i}tre-Robertson-Walker (FLRW) metric: 
\begin{equation}\label{RW1}
ds^{2}= -dt^{2}+a(t)^{2}\delta_{ij}dx^{i}dx^{j}\,,
\end{equation}
where $t$ is the cosmological time, and $a=a(t)$ is the scale factor. For a co-moving observer, the components of the relativistic velocity are: $U^{0} = -1 \,, U^{1}=U^{2}=U^{3} = 0$. Then, tensor and scalar sectors will be presented in the following segments. 
\subsubsection{Tensor back-reaction dynamics}\label{tensor_sector}
The tensor sector is represented by the evolution of $\delta\Psi_{\alpha\beta}$. Due to the symmetry of the waves we have $\delta\Psi_{\alpha\beta}=\delta\Psi_{\beta\alpha}$, and the relevant components will be $\delta\Psi_{ij}$, which for the metric (eq.~(\ref{RW1})) the d'Alembertian operator yields:
\begin{equation}\label{Box-deltaPSI}
\Box \delta\Psi_{ij} = -\delta\ddot{\Psi}_{ij} +\frac{\nabla^{2}}{a^{2}}\delta\Psi_{ij} +H\delta\dot{\Psi}_{ij} + 2\left(\dot{H}+2H^{2}\right)\delta\Psi_{ij} \,,    
\end{equation}
where $H=\dot{a}/a$ is the expansion rate or Hubble parameter. Therefore the dynamics of $\delta\Psi_{ij}$ obeys the equation:
\begin{equation}\label{deltaPSI}
-\delta\ddot{\Psi}_{ij} +\frac{\nabla^{2}}{a^{2}}\delta\Psi_{ij} +H\delta\dot{\Psi}_{ij} + \left(2\dot{H}+4H^{2}\right)\delta\Psi_{ij} = -\frac{6b\Lambda}{M_{P}}\delta\dot{\Psi}_{ij} + \frac{24b\Lambda H}{M_{p}}\delta\Psi_{ij}. 
\end{equation}
We can consider a Fourier expansion for $\delta\Psi_{ij}(t,\underline{x})$, which propagates in an arbitrary spatial direction, which we can make coincident with $\hat{z}$. When sources are considered, there are three transversal modes of the SWs: $+, \times$, and $\textgoth{b}$, where the $\textgoth{b}$-mode denotes the breathing of the wave. In other words, these modes take into account the transversal amplitude of the wave's oscillations. Therefore, the possible transversal components of the wave will be $xx$, $xy$, $yx$, and $yy$, which we shall denote with the letters $m,n$:
\begin{equation}\label{fourier_tensor_expansion}
\delta\Psi_{i\,j}(t,\underline{x}) = \frac{1}{(2\pi)^{2/3}}\sum_{A=+ , \times, \textgoth{b}}\int d^{3}k\, ^{(A)}E_{i j}\left[^{(A)}B_{k}e^{i\underline{k}\cdot\hat{z}}Y_{k}(t) + ^{(A)}B_{k}^{\dagger}e^{-i\underline{k}\cdot\hat{z}}Y^{*}_{k}(t)\right] \,, 
\end{equation}
where $A$ takes into account the degree of freedom of polarisations $+, \times$, and $\textgoth{b}$; then $^{(A)}E_{i j}$ is the polarisation tensor. Here $^{(A)}B_{k}^{\dagger}$ and $^{(A)}B_{k}$ are the creation and the annihilation operators, respectively, for a given (A)–polarisation. The transversal plane $xy$, can be generated by the orthogonal vectors $\underline{u}=(p(t),0)$ and $\underline{v}=(0,q(t))$. Therefore, the components for the polarisation tensor $^{(A)}E_{m n}$, generated by $\underline{u}$ and $\underline{v}$, are:
\begin{equation}
^{(+)}E_{ij}=u_{i}u_{j} - v_{i}v_{j} \,,\quad ^{(\times)}E_{ij}=u_{i}v_{j} + v_{i}u_{j} \,,\quad ^{(\textgoth{b})}E_{ij}=u_{i}u_{j} + v_{i}v_{j} \,,    
\end{equation}
that take the explicit form:
\begin{equation}
^{(+)}E_{ij}= 
\left(
\begin{array}{cc}
p^{2}  & 0      \\
0  &  -q^{2}    
\end{array}
\right)
\,,\qquad 
^{(\times)}E_{ij}= 
\left(
\begin{array}{cc}
0  & pq      \\
pq  &  0    
\end{array}
\right) \,,
\qquad 
^{(\textgoth{b})}E_{ij}= 
\left(
\begin{array}{cc}
p^{2}  & 0      \\
0  &  q^{2}    
\end{array}
\right) \,.
\end{equation}
The modes $Y_{k}$ in eq. \eqref{fourier_tensor_expansion} are described by the differential equation:
\begin{equation}\label{Y_k} 
-\ddot{Y}_{k} + \left(H+\frac{6b\Lambda}{M_{P}}\right)\dot{Y}_{k} + \left[2\dot{H}+4H^{2} - \frac{24b\Lambda H}{M_{P}} - \frac{k^{2}}{a^{2}}\right]Y_{k} = 0 \,, 
\end{equation}
which results from the substitution of eq.~(\ref{fourier_tensor_expansion}) into eq.~(\ref{deltaPSI}). Moreover, given the quantum nature of the fields and their commutation relations, the condition for the tensor modes satisfy:
\begin{equation}\label{norm_cond_Yk}
Y_{k}\dot{Y}_{k}^{*}-\dot{Y}_{k}Y_{k}^{*}=\frac{i}{a^{3}} \,.
\end{equation}
In Fourier space, tensor perturbations can be expressed as the superposition of three polarization modes: $+\,,\times\,,\textgoth{b}$. Each polarisation has a tensor power spectrum, and given that these modes are uncorrelated, the power spectrum of the SWs for all components is straightforwardly computed (see appendix \ref{appendix_a}), yielding:
\begin{equation}\label{tensor-power-spetrum}
P_{Y} = \frac{6 k^{3}}{\pi^{2}}\left|Y_{k}\right|^{2} \,.
\end{equation}
\subsubsection{Scalar back-reaction dynamics}\label{sigma-D}
The scalar sector is described by the dynamics of the scalar back-reaction field $\sigma$ (from eq.~(\ref{equation_sigma})), which obeys the equation: 
\begin{equation}\label{sigma-dynamics}
\ddot{\sigma}+3H\dot{\sigma} - \frac{\nabla^{2}}{a^{2}}\sigma = \frac{6b\,\Lambda}{M_{p}}\dot{\sigma} \,. 
\end{equation}
This field also can be expanded in terms of Fourier series, on the $k$-space as follows:
\begin{equation}\label{fourier_scalar_expansion}
\sigma(t,\underline{x}) = \frac{1}{(2\pi)^{2/3}}\int d^{3}k\, \left[B_{k}e^{i\underline{k}\cdot\hat{z}}Z_{k}(t) + B_{k}^{\dagger}e^{-i\underline{k}\cdot\hat{z}}Z^{*}_{k}(t)\right] \,,    
\end{equation}
where $B_{k}^{\dagger}$ and $B_{k}$ are the creation and the annihilation operators, respectively. And the modes $Z_{k}$ obey the equation:
\begin{equation}\label{Zk}
\ddot{Z}_{k}+\left(3H-\frac{6b\,\Lambda}{M_{p}}\right)\dot{Z}_{k} + \frac{k^{2}}{a^{2}}Z_{k} = 0 \,,
\end{equation}
which comes from the substitution of eq.~(\ref{fourier_scalar_expansion}) into eq.~(\ref{sigma-dynamics}). Once again, given the quantum nature of the fields and their commutation relations, the condition for the scalar modes satisfy:
\begin{equation}\label{norm_cond_Zk}
Z_{k}\dot{Z}_{k}^{*}-\dot{Z}_{k}Z_{k}^{*}=\frac{i}{a^{3}} \,.
\end{equation}
The power spectrum of the scalar modes becomes:
\begin{equation}\label{scalar-power-spetrum}
P_{Z} = \frac{k^{3}}{2\pi^{2}}\left|Z_{k}\right|^{2} \,.
\end{equation}
%


%
%
\section{Back-reaction effects in a single field inflation (or CI) scenario}\label{SW-CI}
In this section, we will study the generation of SWs for an early universe inflation epoch described by a scalar field $\phi$. The Lagrangian density of a minimally-coupled scalar field is given by: 
\begin{equation}\label{lagrangian-phi}
\mathcal{L}_{\phi}= - \frac{1}{2}\partial_{\mu}\phi\partial^{\mu}\phi-V(\phi)\,.
\end{equation}
This is the usual \emph{kinetic-potential energy} form. Thus, the energy-momentum tensor of the inflaton field is:
\begin{equation}\label{energy-momentum-phi}
T_{\mu\nu} = \partial_{\mu}\phi\partial_{\nu}\phi - g_{\mu\nu}\left(\frac{1}{2}\partial_{\alpha}\phi\partial^{\alpha}\phi + V(\phi) \right) \,.
\end{equation}
The cosmological scenario is described by a homogeneous scalar field (inflaton) $\phi(t)$ that carries most of the energy of the universe, hence any other matter content must be subdominant. Then in the cosmological case the universe inflates as the field is rolling down the hill. The dynamic equations are:
\begin{eqnarray} 
&& H^{2} = \frac{\rho_{\phi}}{3M_{P}^{2}} + \frac{\Lambda}{3} \,,\label{CI-equations1}\\
&& \ddot{\phi}+3H\dot{\phi}+V_{\phi} = 0 \,, \label{CI-equations2}
\end{eqnarray}
where $\rho_{\phi}=\dot{\phi}^{2}/2+V(\phi)$ is the energy density of the homogeneous scalar inflaton, $V_{\phi}$ is the derivative of the potential energy with respect to the field and $H=\dot{a}/a$ is the expansion rate or Hubble parameter. The first expression is called Friedmann constraint, the second one is the scalar field equation Klein-Gordon or the energy-momentum conservation equation. As we mentioned before, inflation requires for instance that the density $\rho_{r}$(radiation)$\ll \rho_{\phi}$ in order to generate the abrupt expansion; however, any other present subdominant component might play an important role at the end of inflation. Indeed, this is the study case for the next section in this paper. 
The amount by which the universe inflates is measured as the number of e-foldings $N_{e}$, and since the size of the expansion is expected to be an enormous quantity, it is useful to compute it in terms of the logarithm of the ratio of the scale factor between the end of inflation and a time $t$ during inflation defined by:
\begin{equation}\label{Ne-phi}
N_{e} \equiv \ln\left(\frac{a(t_{end})}{a(t)}\right)=\int_{t}^{t_{end}}dtH \,.
\end{equation}
Typically between 40 and 60 e-foldings of observable inflation\footnote{CMB fluctuations, from the size of the observable universe down to the size of galaxies, were generated during $\sim$10 e-folds about 60 e-folds before the end of inflation.} are large enough to solve the horizon and flatness problems. Finally, inflation last while the slow-roll parameter $\epsilon_{H}=-\dot{H}/H^{2}<1$, where $\dot{H} = -\dot{\phi}^{2}/2/M_{P}^{2}$, so it ends when $\epsilon_{H}=1$. Moreover, an inflationary scenario also requires a flat potential. This condition is measured by the $\eta_{H}=-\ddot{\phi}/(H\dot{\phi})$ parameter. Where a particular potential preserves flatness having $|\eta_{H}|<1$. Finally, to describe the generation of SWs we introduce a chaotic quartic potential:
\begin{equation} 
V(\phi) = \frac{\lambda}{4}\phi^{4} \,,
\label{quarticpotential}
\end{equation}
albeit this potential is already ruled out from current observations \cite{akrami:2018b}, we are more interested of showing a comprehensive study of tensor and scalar modes due to the back-reaction effects. We present the outcome of the CI framework illustrated by a chaotic quartic potential for about $N_{e}\simeq 60$. We take $\lambda=2.2315\times 10^{-13}$ and the initial value of the inflaton $\phi_{0}=22.01 \, M_{P}$. Inflation lasts $N_{e}=60.6$. Also, we consider the pivot scale~{\footnote{In fact, the Planck collaboration adopts a default pivot scale $k_{*} = 0.05 \,\rm Mpc^{-1}$, since this value corresponds to previous Planck releases; however, they also quote another scale $k_{0} = 0.002 \,\rm Mpc^{-1}$ in order to facilitate comparison with earlier primordial tensor-mode constraints. So we consider the later value.}} $k_{0}\simeq 0.002 \,\rm Mpc^{-1}$ (at $N_{e}^{*}=8.12$) that corresponds to the CMB observation by Planck Legacy pivot value~\cite{akrami:2018b, Planck:2018jri}. We solve numerically eqs.~(\ref{Y_k},\ref{Zk}), the variables $\psi_{k}=a Z_{k}$ and $\alpha_{k}=a Y_{k}$ are introduced to set the initial conditions (see Appendix~\ref{appendix_b} for the initial conditions setting). We assume that $\mathbf{6 b/M_{P}=m\,H/\Lambda}$ (here $\mathbf{m} \epsilon\,\mathbb{R}$), since $H$ is nearly constant until nearly before inflation ends. Also we consider the regime $k\gg aH$, when the scales are well inside the horizon. The quantum nature of the fields allow us to set the values of $m$ where the normalisation conditions are satisfied (eqs.~(\ref{norm_cond_Yk},\ref{norm_cond_Zk})). We find that for the scalar sector $m$ must be zero or an even integer (positive or negative see fig.~\ref{fig:scalar_norm_condition}). On the other hand, in the tensor sector we have, for example, with $m=-2$ the constraint $Y_{k}\dot{Y}_{k}^{*}-\dot{Y}_{k}Y_{k}^{*}=ia^{-3}$ is satisfied (see fig.~\ref{fig:tensor_norm_condition}). 

\begin{figure}[htbp] 
\includegraphics[scale=0.95]{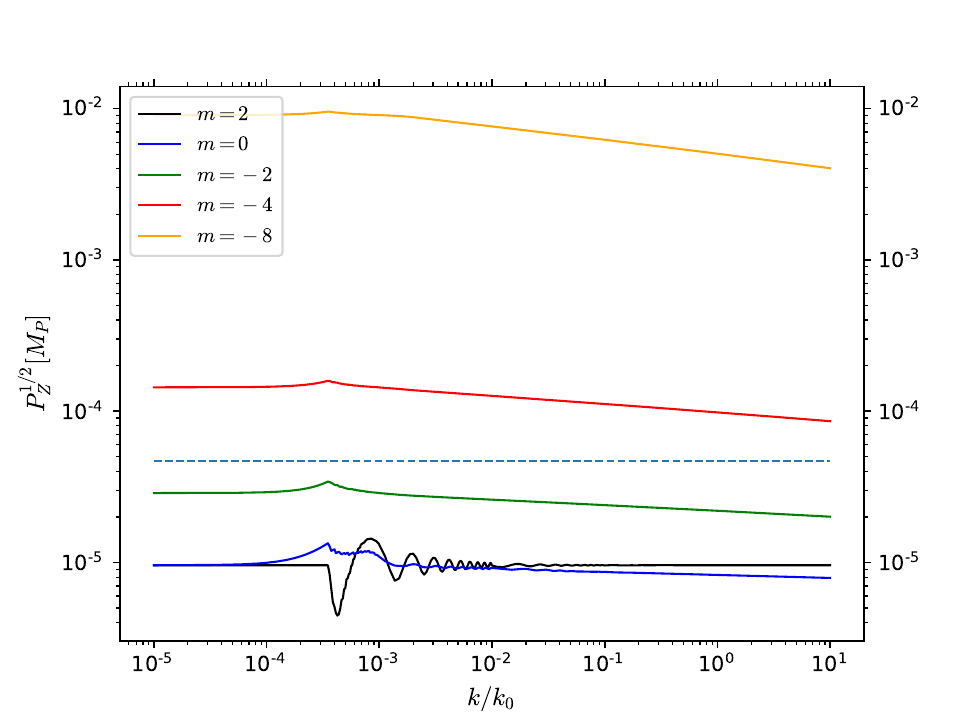}
\caption{Behaviour of the square root of the power spectrum of the scalar modes $P_{Z}^{1/2}$ versus $k/k_{0}$ (here $k_{0}\simeq 0.002 \rm Mpc^{-1}$), described by a quartic potential for $>60$ e-folds of inflation. We show the outcome for five different $m$'s: $m=2$ black; $m=0$ blue; $m=-2$ green; $m=-4$ red; and $m=-8$ orange. As reference, the dashed light blue horizontal line represents the observational value of the primordial curvature power spectrum $P_{\mathcal{R}_{\phi}}^{1/2}\simeq 4.7\times 10^{-5}$ \cite{Planck:2018jri}, where only the outcomes with $m=-2$ (green) and $m=-4$ (red) are close to this value. The smaller $m$ is the larger the value of $P_{Z}^{1/2}$. However, when $m=2$ then $P_{Z}^{1/2}$ oscillates from about $k/k_{0}\simeq 3\times 10^{-4}$ to $k/k_{0}\simeq 0.1$ and from there becomes constant. And the rest of the upshots slowly decrease for larger $k/k_{0}$.}\label{fig:various_PZ}
\end{figure}

The resulting numerical analysis is presented, for instance, in fig.~\ref{fig:various_PZ} where we have plotted the square root of the power spectrum of the scalar modes $P_{Z}^{1/2}$ versus $k/k_{0}$; we show the outcome for five different $m$'s: $m=2$ black; $m=0$ blue; $m=-2$ green; $m=-4$ red; and $m=-8$ orange. The smaller $m$ is the larger the value of $P_{Z}^{1/2}$. However, when $m=2$ then $P_{Z}^{1/2}$ oscillates from about $k/k_{0}\simeq 3\times 10^{-4}$ to $k/k_{0}\simeq 0.1$ and from there becomes constant, and the rest of the upshots slowly decrease for larger $k/k_{0}$. Moreover, as reference, we put the observational value of the primordial curvature power spectrum $P_{\mathcal{R}_{\phi}}^{1/2}\simeq 4.7\times 10^{-5}$ \cite{Planck:2018jri} (the dashed light blue horizontal line) to contrast all results. Only the outcomes with $m=-2$ (green) and $m=-4$ (red) are close to the Planck value $P_{\mathcal{R}_{\phi}}^{1/2}\simeq 4.7\times 10^{-5}$, the rest of them are either much larger or much smaller than $P_{\mathcal{R}_{\phi}}^{1/2}$. 

\begin{figure}[htbp] 
\includegraphics[scale=0.95]{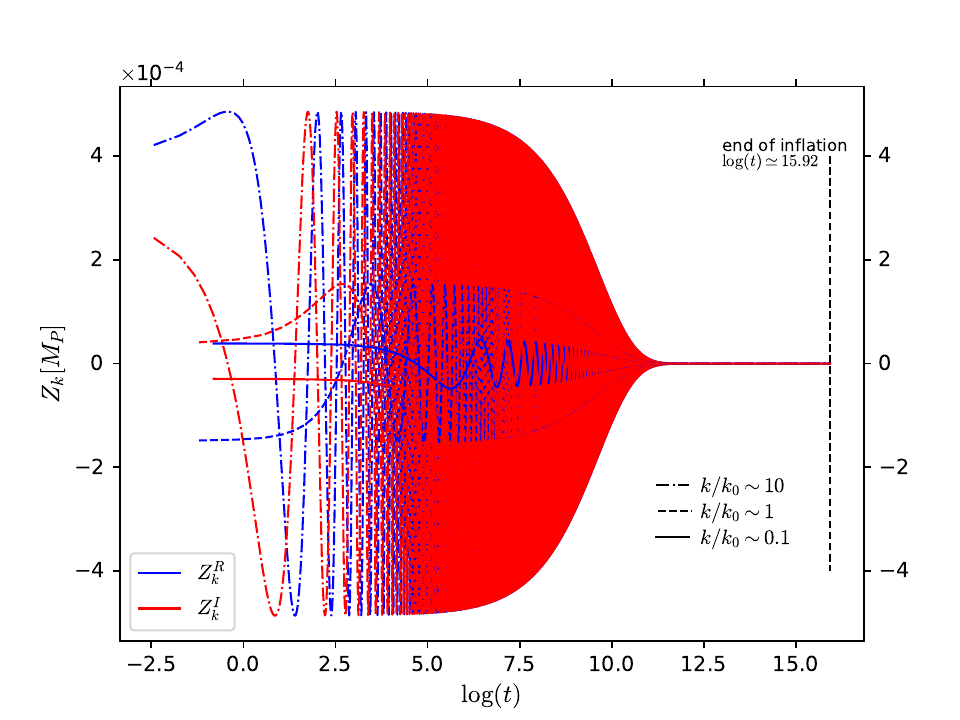}
\caption{Behaviour of scalar modes $Z_{k}^{\mathbb{R}}$ (blue) and $Z_{k}^{\mathbb{I}}$ (red) versus $\log(t)$, described by a quartic potential for $>60$ e-folds of inflation with $m=-2$. We show the outcome for three different wave numbers $k/k_{0}$ (here $k_{0}\simeq 0.002 \rm Mpc^{-1}$): $k/k_{0}=0.1$ solid lines; $k/k_{0}=1$ dashed ones; and $k/k_{0}=10$ are described by dash-dotted lines. Note that all modes become constant at (or right after) horizon crossing $k\simeq aH$ ($\log(t)\gtrsim 12.5$). Note that the larger $k/k_{0}$ the $Z_{k}$'s amplitudes also increase.}\label{fig:m_-2_Z}
\end{figure}

\begin{figure}[htbp] 
\includegraphics[scale=0.95]{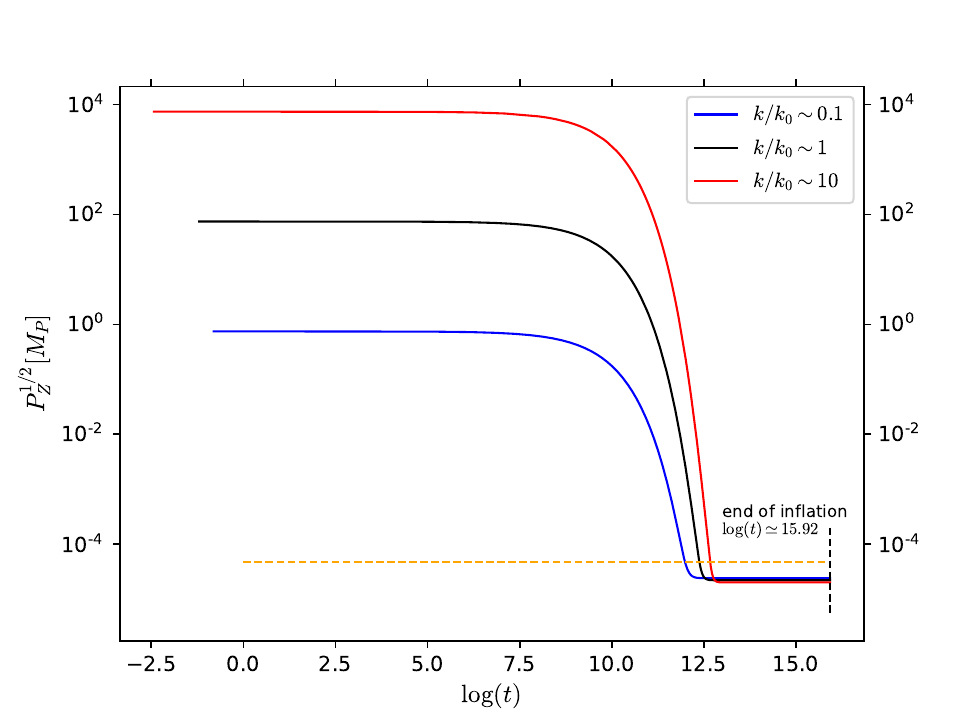}
\caption{Behaviour of the square root of the power spectrum of the scalar modes $P_{Z}^{1/2}$ versus $\log(t)$, described by a quartic potential for $>60$ e-folds of inflation with $m=-2$. We show the outcome for three different wave numbers $k/k_{0}$ (here $k_{0}\simeq 0.002 \rm Mpc^{-1}$): $k/k_{0}=0.1$ blue; $k/k_{0}=1$ black; and $k/k_{0}=10$ red. As reference, the dashed orange horizontal line represents the observational value of the primordial curvature power spectrum $P_{\mathcal{R}_{\phi}}^{1/2}\simeq 4.7\times 10^{-5}$ \cite{Planck:2018jri}. Note that all modes become constant at (or right after) horizon crossing $k\simeq aH$ ($\log(t)\gtrsim 12.5$), and they asymptotically converge to a smaller value than $P_{\mathcal{R}_{\phi}}^{1/2}\simeq 4.7\times 10^{-5}$.}\label{fig:m_-2_PZ}
\end{figure}

In addition, we present a particular example with $m=-2$ (see figs.~\ref{fig:m_-2_Z} and \ref{fig:m_-2_PZ}) for three different wave numbers $k/k_{0}$: $k/k_{0}=0.1$ solid lines; $k/k_{0}=1$ dashed ones; and $k/k_{0}=10$ are described by dash-dotted lines. The scalar modes $Z_{k}^{\mathbb{R}}$ (blue) and $Z_{k}^{\mathbb{I}}$ (red) oscillate within the horizon, and they become constant at (or right after) horizon crossing $k\simeq aH$ ($\log(t)\gtrsim 12.5$). Note that the larger $k/k_{0}$ the $Z_{k}$'s amplitudes increase as well. Also, in fig.~\ref{fig:m_-2_PZ} we show the behaviour of the square root of the power spectrum of the scalar modes $P_{Z}^{1/2}$ versus $\log(t)$, and this time the dashed orange horizontal line represents the observational value of the primordial curvature power spectrum $P_{\mathcal{R}_{\phi}}^{1/2}\simeq 4.7\times 10^{-5}$ \cite{Planck:2018jri}. We confirm that all modes become constant at (or right after) horizon crossing, and they asymptotically converge to a smaller value than $P_{\mathcal{R}_{\phi}}^{1/2}\simeq 4.7\times 10^{-5}$.

\begin{figure}[htbp] 
\includegraphics[scale=0.95]{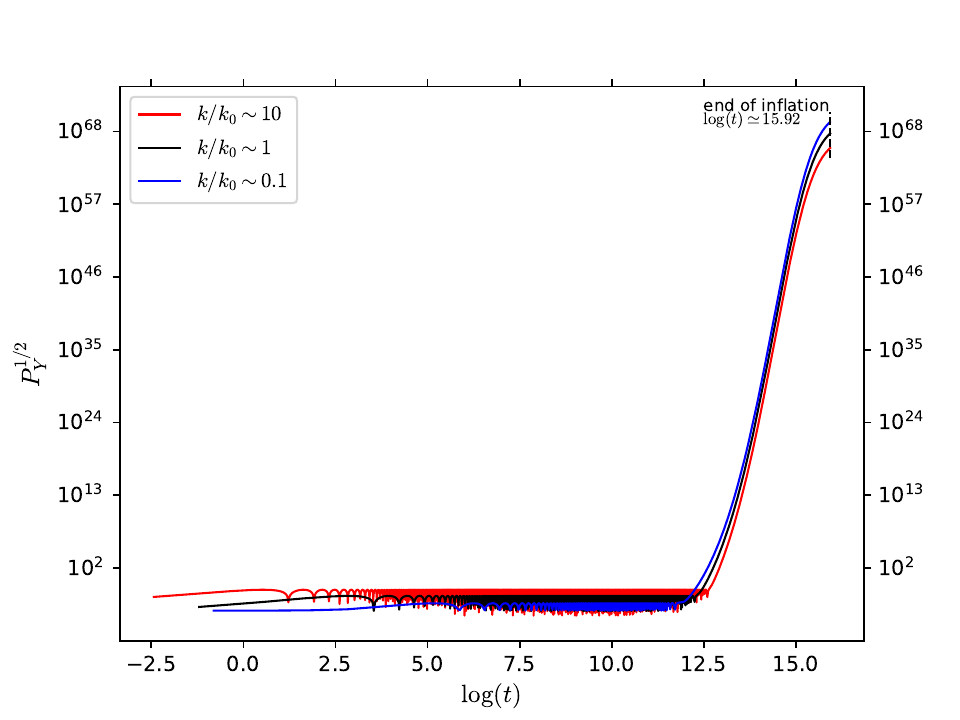}
\caption{Behaviour of the square root of the power spectrum of the tensor modes $P_{Y}^{1/2}$ versus $\log(t)$, described by a quartic potential for $>60$ e-folds of inflation with $m=-2$. We show the outcome for three different wave numbers $k/k_{0}$ (here $k_{0}\simeq 0.002 \rm Mpc^{-1}$): $k/k_{0}=0.1$ blue; $k/k_{0}=1$ black; and $k/k_{0}=10$ red. Note that all modes oscillate inside the horizon but they abruptly grow just as they cross it. This upshot hinders any probable observational hint or signal.}\label{fig:tensor_m_-2_PY}
\end{figure}

On the other hand, in fig.~\ref{fig:tensor_m_-2_PY} we present the evolution of the square root of the power spectrum of the tensor modes $P_{Y}^{1/2}$ versus $\log(t)$. The tensor sector, however, shows an irregular journey due to its abruptly growth just as it crosses the horizon. This might indicate that tensor modes must interact nearly the entire duration of inflation, therefore, any relevant scale must leave the horizon just about inflation ends, i.e., at $k^{(tensor)}\simeq H_{end}a_{end}$. However, our main methodology is not longer valid at this regime, so we must abandon this prescription and develop an improved one in order to present a reliable explanation of any probable observational hint or signal.

%
\section{Back-reaction effects in Warm Inflation (WI)}\label{SW-WI}
In this section, we study the generation of spacetimes ripples within a WI framework. To begin with, WI brings radiation into the evolution of inflation. Thus, a thermal bath imbibes the inflaton field, modifying the dynamics both at the background and the perturbation level. At first the main input is to introduce a dissipative term $\Upsilon\dot{\phi}$ in the inflaton evolution equation, as a source of radiation production \cite{Berera:1995wh,Berera:1995ie,Berera:1996fm,Berera:2008ar}. To be able to describe this mechanism, all macroscopic motion must be slow relative to the relevant microscopic time scales, then the macroscopic dynamics can be treated adiabatically. To ensure the aforementioned near-equilibrium process, one must satisfy the adiabatic condition, namely $\Gamma/H>1$, where $\Gamma$ is the decay width, which quantifies the microscopic interactions; and $H$ corresponds to a macroscopic measurement. Posteriorly, a noise force term was included to also drive the inflaton fluctuations, with a fluctuation-dissipation theorem uniquely specifying the inflaton fluctuations \cite{Berera:1995wh,Berera:1995ie,Berera:1996fm,Berera:2008ar}. Thus, we will start with the WI background evolution equations for the inflaton-radiation system, given by: 
\begin{eqnarray}
\ddot{\phi}+(3H+\Upsilon)\dot{\phi}+ V_{\phi} &=& 0 \,, \nonumber\\
\dot{\rho}_{r}+4H\rho_{r} &=& \Upsilon\dot{\phi}^{2} \,, \label{background_eqs}
\end{eqnarray}
where $\Upsilon=\Upsilon(T,\phi)$ is the dissipative coefficient{\footnote{Indeed, $\Upsilon$ can depend on either temperature $T$ or $\phi$ or both. For instance, a particular supersymmetric model yields a dissipative coefficient $\Upsilon \propto T^{3}/\phi^{2}$ \cite{Bastero-Gil:2009sdq}.}} in the leading adiabatic approximation, and it is computed from first principles provided the relevant interactions between the scalar field and the thermalised degrees of freedom (dof); $\rho_{r}=\pi^{2}g_{*}T^{4}/30$, $g_{*}$ being the effective no. of light dof. And once again dots correspond to time derivatives, $V_{\phi}$ is the derivative of the potential energy with respect to the field, and $H$ is the Hubble parameter, given by the Friedmann equation for a flat FRW universe:
\begin{equation}\label{Hubble}
H^{2}=\frac{\rho}{3M_P^{2}} + + \frac{\Lambda}{3} \,,
\end{equation}
where $\rho=\rho_{\phi}+\rho_{r}$ is the total energy density, with $\rho_{\phi}=\dot{\phi}^{2}/2+V(\phi)$. Also the field pressure and the equation of state for the radiation degrees of freedom are $p_{\phi}=\dot{\phi}^{2}/2-V(\phi)$ and $p_{r}=\rho_{r}/3$ respectively. The dissipation coefficient is given by \cite{Bastero-Gil:2018yen}:
\begin{equation}
\Upsilon = C_{\phi}\phi^{n}\,, \quad n\geq 1 \,,  
\end{equation}
where this expression comes from the dynamics of a supersymmetric distributed mass (DM) model. This particular scheme arises from a general form of an effective $\rm N=1$ global Supersymmetry (SUSY) theory version of the DM model with chiral superfields $\Phi$, $X_{i}$ and $W_{i}$ \cite{Hall:2004zr, Bastero-Gil:2009sdq, Bastero-Gil:2010dgy, Bastero-Gil:2012akf}. The chiral superfields $\Phi$, $X_{i}$ and $W_{i}$ have (scalar, fermion) components ($\phi$,$\psi_{\phi}$), ($\chi_{i}$,$\psi_{\chi_{i}}$) and ($\sigma_{i}$,$\psi_{\sigma_{i}}$), respectively. Thus, $g_{*}= 1 + 15 N_M/4$, $N_M$ being the no. of bosonic $\chi_i$ (fermionic $\psi_i$) light degrees of freedom at horizon crossing. There are more consistent with microscopic derivations of a the temperature independent $\Upsilon$ (see, e.g., refs.\cite{deOliveira:1997jt, delCampo:2007cy, delCampo:2010by, Ramos:2001zw, Ramos:2013nsa}). In addition, in the DM model, the average decay width $\Gamma \propto T$ \cite{Bastero-Gil:2018yen}. During inflation the motion of the inflaton field has to be overdamped in order to end the accelerated expansion, and indeed this can be achieved due to either the Hubble rate, as the CI case, or an extra friction $\Upsilon$ term, or the interaction of both components. We can quantify this competition by defining the dissipative ratio $Q=\Upsilon/(3H)$. According to the ratio $Q$ we will have distinct cases: for $Q<1$, this is called weak dissipative warm inflation (WDWI); and when $Q\geq 1$ we are in strong dissipative warm inflation (SDWI). Importantly, given the presence of radiation during inflation, the potential acquires thermal corrections, and indeed they become relevant at the background and fluctuation levels, since they adjust the slow-roll parameter $\epsilon_{H}$. However, in the scenario with a quartic chaotic potential eq. \eqref{quarticpotential} within the SDWI, such thermal corrections give a negligible contribution to both the effective potential and its first derivative in the perturbative regime \cite{Bastero-Gil:2018yen}. Hence, we will not consider these thermal corrections. Moreover, in order to describe a near-equilibrium process, one must satisfy thermalisation, that is the adiabatic condition $\Gamma/H>1$, which translates roughly into $T/H>1$ since $\Gamma \propto T$ \cite{Bastero-Gil:2018yen}. To describe the generation of SWs, we anew introduce a chaotic quartic potential: $V(\phi) = \lambda\phi^{4}/4$, and we utilise a linear dissipative coefficient: 
\begin{equation}\label{linear_dc}
\Upsilon = C_{\phi}\phi \,. 
\end{equation}
To understand the use of above constituents, let us refer to certain published features. First, this example works properly without adding thermal contributions to the potential, provided the outcome comes from a SDWI \cite{Bastero-Gil:2018yen}, which is this case of study. Second, the adiabatic condition $\Gamma/H>1$ or $T/H>1$ is easily achieved in the SDWI; so there is no need to examine this thoroughly. Third, even though we mentioned before that above potential within CI is already ruled out from current observations \cite{akrami:2018b}, when including a WI scenario, this model lies within Planck observational window \cite{akrami:2018b}. Note that from \cite{Bastero-Gil:2018yen} authors obtained that the scalar spectral index $n_{s}-1\simeq d\ln P_{\mathcal{R}}/dN_{e}$ gets values of $n_{s}(N_{e}=50)=0.9567$ and $n_{s}(N_{e}=60)=0.9637$, where at 60 e-folds this scalar spectral index agrees outstandingly with Planck data: $n_{s}=0.9649\pm 0.0042$ at $68\%$ CL \cite{akrami:2018b}; and for the tensor to scalar ratio $r=P_{t}/P_{\mathcal{R}}$, where $P_{t}=2H^{2}/\pi^{2}/M_{P}^{2}$ is the tensor power spectrum of primordial GWs that corresponds to the tensor part of the metric perturbation{\footnote{With the purpose of not confusing the reader, these tensor modes come from the tensor part of the metric perturbation, which are enhanced due to the sudden expansion at the very early universe (for more details see \cite{Lyth:2009zz,Baumann:2009ds,Dodelson:2003ft}). And these do not refer to the ones that we are indeed analysing in this paper.}}, they obtained $10^{-9}< r\lesssim 10^{-4}$ at $10\leq Q_{*}\leq 1000$, for $N_M=O(10-100)$, having the bigger suppression at the largest $Q_{*}$; thus the tensor-to-scalar ratio is highly suppressed by dissipation. Last but not the least, two swampland criteria, relevant for inflationary theories, have been intensively discussed \cite{Obied:2018sgi,Agrawal:2018own}, $|\Delta\phi|/M_{P}<\Delta$ and $M_{P}|V_{\phi}|/V>c$, provided that $V>0$, where $\{\Delta,c\}\sim\mathcal{O}(1)$. And these criteria are very consistent with a WI paradigm described by a DM model \cite{Bastero-Gil:2018yen}. In fact, it was noted that inherently the dissipative feature in WI makes it amenable for consistency with swampland criteria, as already noted in the literature \cite{Das:2018hqy,Das:2018rpg,Motaharfar:2018zyb,Yi:2018dhl,Lin:2018edm}.

\begin{figure}[htbp] 
\includegraphics[scale=0.95]{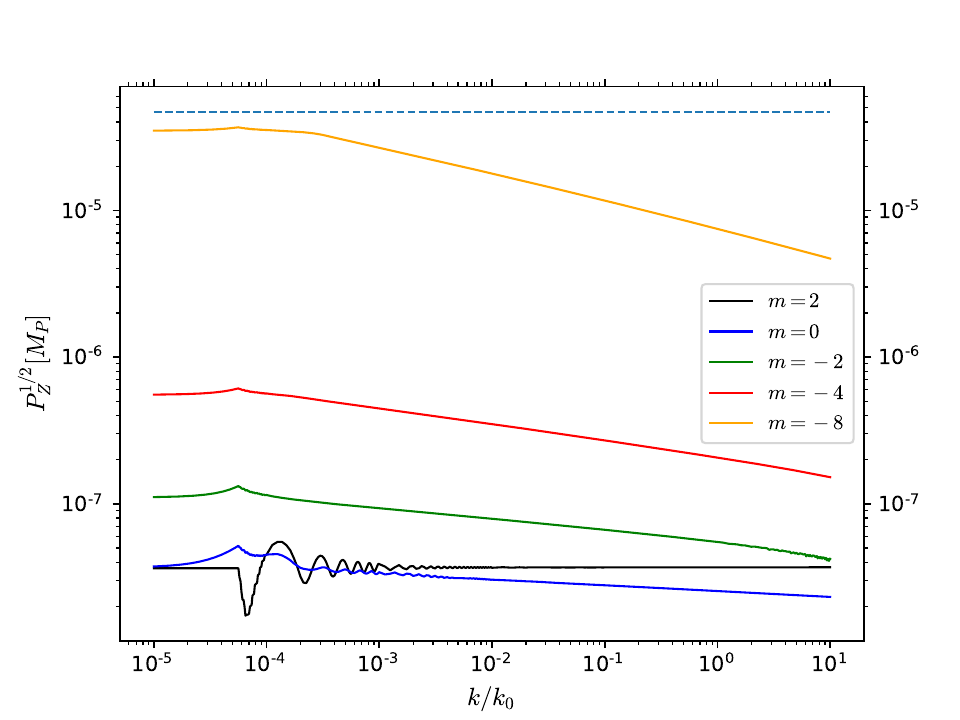}
\caption{Behaviour of the square root of the power spectrum of the scalar modes $P_{Z}^{1/2}$ versus $k/k_{0}$ (here $k_{0}\simeq 0.002 \rm Mpc^{-1}$), described by a quartic potential and a linear dissipative coefficient for $>60$ e-folds of inflation. We show the outcome for five different $m$'s: $m=2$ black; $m=0$ blue; $m=-2$ green; $m=-4$ red; and $m=-8$ orange. As reference, the dashed light blue horizontal line represents the observational value of the primordial curvature power spectrum $P_{\mathcal{R}_{\phi}}^{1/2}\simeq 4.7\times 10^{-5}$ \cite{Planck:2018jri}. The smaller $m$ is the larger the value of $P_{Z}^{1/2}$. However, when $m=2$ then $P_{Z}^{1/2}$ oscillates from about $k/k_{0}\simeq 6\times 10^{-5}$ to $k/k_{0}\simeq 0.01$ and from there becomes constant. The rest of the upshots slowly decrease for larger $k/k_{0}$ and their values are smaller than the CI ones.}\label{fig:WI_scalar_modes_varios}
\end{figure}

\begin{figure}[htbp] 
\includegraphics[scale=0.95]{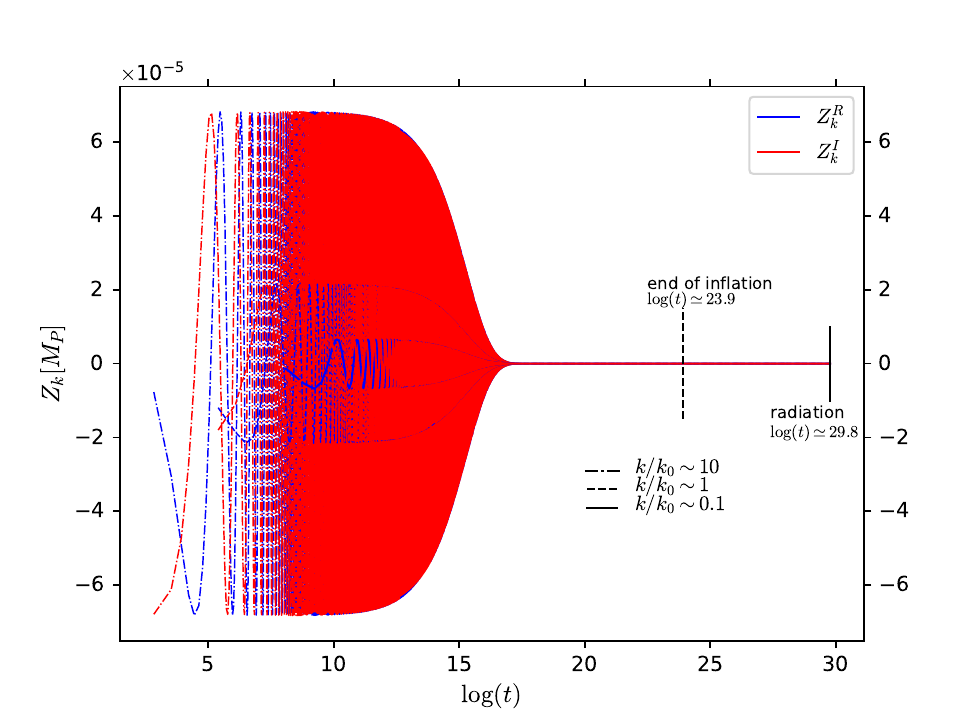}
\caption{Behaviour of scalar modes $Z_{k}^{\mathbb{R}}$ (blue) and $Z_{k}^{\mathbb{I}}$ (red) versus $\log(t)$, described by a quartic potential and a linear dissipative coefficient for $>60$ e-folds of inflation and $m=-2$. We show the outcome for three different wave numbers $k/k_{0}$ (here $k_{0}\simeq 0.002 \rm Mpc^{-1}$): $k/k_{0}=0.1$ solid lines; $k/k_{0}=1$ dashed ones; and $k/k_{0}=10$ are described by dash-dotted lines. Note that both modes become constant at (or right after) horizon crossing $k\simeq aH$ ($\log(t)\gtrsim 18.5$) and they remain so when radiation starts to dominate. There is approximately an order of magnitude reduction in amplitude with respect to the CI scenario.}\label{fig:WI_m-2_Z}
\end{figure}

\begin{figure}[htbp] 
\includegraphics[scale=0.95]{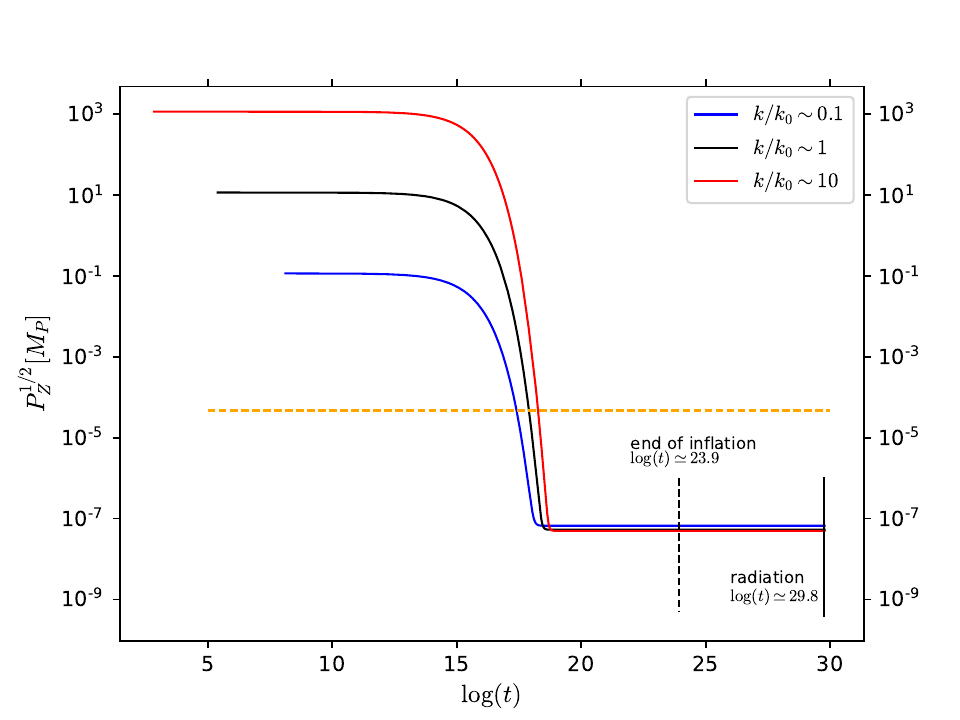}
\caption{Behaviour of the square root of the power spectrum of the scalar modes $P_{Z}^{1/2}$ versus $\log(t)$, described by a quartic potential and a linear dissipative coefficient for $>60$ e-folds of inflation with $m=-2$. We show the outcome for three different wave numbers $k/k_{0}$ (here $k_{0}\simeq 0.002 \rm Mpc^{-1}$): $k/k_{0}=0.1$ blue; $k/k_{0}=1$ black; and $k/k_{0}=10$ red. As reference, the dashed orange horizontal line represents the observational value of the primordial curvature power spectrum $P_{\mathcal{R}_{\phi}}^{1/2}\simeq 4.7\times 10^{-5}$ \cite{Planck:2018jri}. Note that all modes become constant at (or right after) horizon crossing $k\simeq aH$ ($\log(t)\gtrsim 18.5$) remaining so when radiation starts to dominate. They asymptotically converge to a much smaller value than the CI one (see fig.~\ref{fig:m_-2_PZ}).}\label{fig:WI_m-2_PZ}
\end{figure}

\begin{figure}[htbp] 
\includegraphics[scale=0.75]{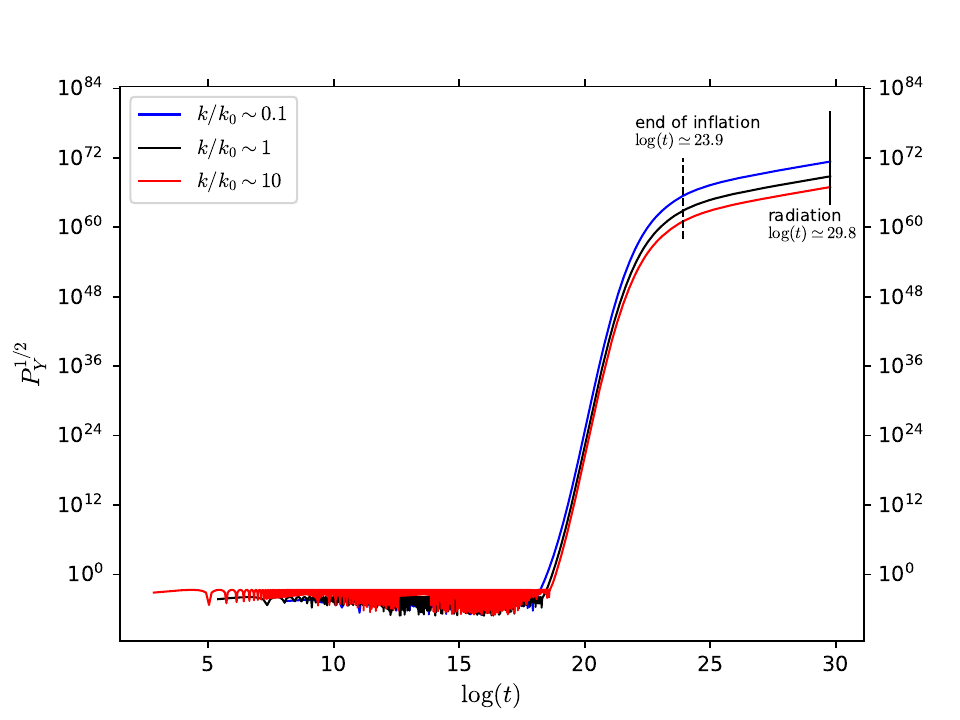}
\caption{Behaviour of the square root of the power spectrum of the tensor modes $P_{Y}^{1/2}$ versus $\log(t)$, described by a quartic potential and a linear dissipative coefficient for $>60$ e-folds of inflation with $m=-2$. We show the outcome for three different wave numbers $k/k_{0}$ (here $k_{0}\simeq 0.002 \rm Mpc^{-1}$): $k/k_{0}=0.1$ blue; $k/k_{0}=1$ black; and $k/k_{0}=10$ red. Note that all modes oscillate inside the horizon but they abruptly grow just as they cross it; however, right after inflation ends, this sudden surge moderates when radiation is about to dominate. This result might indicate that radiation would help to mitigate the exponential rise of $P_{Y}^{1/2}$, allowing us to search for an observational footprint.}\label{fig:WI_tensor_modes}
\end{figure}

Our analysis (see Appendix~\ref{appendix_b} for the initial conditions setting) is presented in the following plots figs.~\ref{fig:WI_scalar_modes_varios}, \ref{fig:WI_m-2_Z}, \ref{fig:WI_m-2_PZ},  and \ref{fig:WI_tensor_modes}. Numerically, we take $C_{\phi}=3.88\times 10^{-6}$, $N_{M}=10$, therefore $g_{*}=77/2$. In addition, the initial value of the inflaton $\phi_{0}=3.455 \, M_{P}$, and $\lambda=4.45\times 10^{-15}$ (which is the normalised value fixed by the CMB value of $P_{\mathcal{R}_{\phi}}^{1/2}\simeq 4.7\times 10^{-5}$ \cite{akrami:2018b}). Inflation lasts $N_{e}=60.22$. Once again we take the pivot scale $k_{0}\simeq 0.002 \,\rm Mpc^{-1}$ (at $N_{e}^{*}=10.15$) that corresponds to the CMB observation by Planck Legacy pivot value~\cite{akrami:2018b, Planck:2018jri}. Given the aforementioned set of input values, we guarantee that the system is within a SDWI regime; where, in fact, the parameter $Q_{*}\simeq 23.0$. And we let evolve the system until radiation starts dominating at $N_{e}^{rad}=64.33$ ($\epsilon_{H}=2$).   

Fig.~\ref{fig:WI_scalar_modes_varios} shows the square root of the power spectrum of the scalar modes $P_{Z}^{1/2}$ versus $k/k_{0}$; we show the outcome for five different $m$'s: $m=2$ black; $m=0$ blue; $m=-2$ green; $m=-4$ red; and $m=-8$ orange. The smaller $m$ is the larger the value of $P_{Z}^{1/2}$. However, when $m=2$ then $P_{Z}^{1/2}$ oscillates from about $k/k_{0}\simeq 6\times 10^{-5}$ to $k/k_{0}\simeq 0.01$ and from there becomes constant, and the rest of the upshots slowly decrease for larger $k/k_{0}$. Once again, as reference, we put the observational value of the primordial curvature power spectrum $P_{\mathcal{R}_{\phi}}^{1/2}\simeq 4.7\times 10^{-5}$ \cite{Planck:2018jri} (the dashed light blue horizontal line) to contrast all results. This time all the outcomes are smaller or much smaller than $P_{\mathcal{R}_{\phi}}^{1/2}$.

Moreover, we present a particular example with $m=-2$ (see figs.~\ref{fig:WI_m-2_Z} and \ref{fig:WI_m-2_PZ}) for three different wave numbers $k/k_{0}$: $k/k_{0}=0.1$ solid lines; $k/k_{0}=1$ dashed ones; and $k/k_{0}=10$ are described by dash-dotted lines. The scalar modes $Z_{k}^{\mathbb{R}}$ (blue) and $Z_{k}^{\mathbb{I}}$ (red) oscillate within the horizon, and they become constant at (or right after) horizon crossing $k\simeq aH$ ($\log(t)\gtrsim 18.5$) and they remain so when radiation starts to dominate. Note that the larger $k/k_{0}$ the $Z_{k}$'s amplitudes increase too. Also, in fig.~\ref{fig:WI_m-2_PZ} we show the behaviour of the square root of the power spectrum of the scalar modes $P_{Z}^{1/2}$ versus $\log(t)$, and this time the dashed orange horizontal line represents the observational value of the primordial curvature power spectrum $P_{\mathcal{R}_{\phi}}^{1/2}\simeq 4.7\times 10^{-5}$ \cite{Planck:2018jri}. We confirm that all modes become constant at (or right after) horizon crossing and remaining so when radiation starts to dominate; besides they asymptotically converge to a much smaller value than $P_{\mathcal{R}_{\phi}}^{1/2}\simeq 4.7\times 10^{-5}$. In general we can notice that radiation reduces the size of the $Z_{k}$'s amplitudes, hence yielding smaller signals of such modes.


Finally in fig.~\ref{fig:WI_tensor_modes} we present the evolution of the square root of the power spectrum of the tensor modes $P_{Y}^{1/2}$ versus $\log(t)$. Once again the tensor sector shows an irregular behaviour at horizon crossing; however, this time due to radiation this enormous growth slows down during the transition between the end of inflation and radiation domination. We must investigate this further provided an improved methodology is presented in future works.     

%
%
\section{Conclusions}\label{conclusions}
The study of early universe cosmology presents an awe-inspiring realm filled with fascinating and unconventional physics. Thus we require more sophisticated tools to accurately describe and understand this intricate framework. This work pursues the objective of studying a scenario in which geometrical scalar back-reaction effects in inflation are considered. We have conducted an investigation into the production of SWs caused by the presence of a closed manifold, along with its geometric boundary terms, within two frameworks: CI and WI. 

In both schemes we present the outcome of a chaotic eq. \eqref{quarticpotential} for about $N_{e}\simeq 60$. Also, we consider the pivot scale $k_{0}\simeq 0.002 \,\rm Mpc^{-1}$ that corresponds to the CMB observation by Planck Legacy pivot value~\cite{akrami:2018b, Planck:2018jri}. In WI we utilise a linear dissipative coefficient $\Upsilon=C_{\phi}\phi$, within a SDWI regime, where the dissipative ratio is $Q_{*}\simeq 23.0$. To implement our numerical analysis the variables $\psi_{k}=a Z_{k}$ and $\alpha_{k}=a Y_{k}$ are introduced to set the initial conditions (see Appendix~\ref{appendix_b} for the initial conditions setting). We assume that $\mathbf{6 b/M_{P}=m\,H/\Lambda}$ (here $m\epsilon\,\mathbb{R}$), since $H$ is nearly constant until nearly before inflation ends. Also we consider the regime $k\gg aH$, when the scales are well inside the horizon. The quantum nature of the fields allow us to set the values of $m$ where the normalisation conditions are satisfied (eqs.~(\ref{norm_cond_Yk}, \ref{norm_cond_Zk})). We find that for the scalar sector $m$ must be zero or an even integer (positive or negative see fig.~\ref{fig:scalar_norm_condition}). On the other hand, in the tensor sector we have, for example, with $m=-2$ the constraint $Y_{k}\dot{Y}_{k}^{*}-\dot{Y}_{k}Y_{k}^{*}=ia^{-3}$ is satisfied (see fig.~\ref{fig:tensor_norm_condition}). 

In the scalar sector the behaviour of the square root of the power spectrum of the scalar modes $P_{Z}^{1/2}$ versus $k/k_{0}$ is presented in fig.~\ref{fig:various_PZ} (CI) and fig. \ref{fig:WI_scalar_modes_varios} (WI). As reference, we put the observational value of the primordial curvature power spectrum $P_{\mathcal{R}_{\phi}}^{1/2}\simeq 4.7\times 10^{-5}$ \cite{Planck:2018jri} (the dashed light blue horizontal line) to contrast all results. In the CI case, only the outcomes with $m=-2$ (green) and $m=-4$ (red) are close to the Planck value, the rest of them are either much larger or much smaller than $P_{\mathcal{R}_{\phi}}^{1/2}$. Contrary to the WI instance where all results are much smaller than $P_{\mathcal{R}_{\phi}}^{1/2}$. On the other hand, we present a particular example with $m=-2$ (see figs.~\ref{fig:m_-2_Z}, \ref{fig:m_-2_PZ}, \ref{fig:WI_m-2_Z}, and \ref{fig:WI_m-2_PZ}) for three different wave numbers $k/k_{0}$: $k/k_{0}=0.1$ solid lines; $k/k_{0}=1$ dashed ones; and $k/k_{0}=10$ are described by dash-dotted lines. The scalar modes $Z_{k}^{\mathbb{R}}$ (blue) and $Z_{k}^{\mathbb{I}}$ (red) oscillate within the horizon, and they become constant at (or right after) horizon crossing $k\simeq aH$ and they remain so when radiation starts to dominate (figs.~\ref{fig:m_-2_Z} and \ref{fig:WI_m-2_Z}). The larger $k/k_{0}$ the $Z_{k}$'s amplitudes increase too. Also, in figs.~\ref{fig:m_-2_PZ} and \ref{fig:WI_m-2_PZ} we show the behaviour of the square root of the power spectrum of the scalar modes $P_{Z}^{1/2}$ versus $\log(t)$, where this time the dashed orange horizontal line represents the observational value of the primordial curvature power spectrum $P_{\mathcal{R}_{\phi}}^{1/2}\simeq 4.7\times 10^{-5}$ \cite{Planck:2018jri}. We confirm that all modes become constant at (or right after) horizon crossing and remaining so when radiation starts to dominate; besides they asymptotically converge to a much smaller value than $P_{\mathcal{R}_{\phi}}^{1/2}\simeq 4.7\times 10^{-5}$. In general we can notice that radiation reduces the size of the $Z_{k}$'s amplitudes, hence yielding smaller signals of such modes. 

The tensor sector in both scenarios shows an irregular journey due to their abruptly growth just as they cross the horizon, see figs.~\ref{fig:tensor_m_-2_PY} and \ref{fig:WI_tensor_modes}. And, in fact, this upshot hinders any probable observational hint or signal. However, in WI due to radiation this enormous growth slows down during the transition between the end of inflation and radiation domination (fig.~\ref{fig:WI_tensor_modes}). We believe this unreliable evolution is mainly due to our approach of setting the initial conditions. First, we assume that oscillation occurs when the scales are well inside the horizon ($k\gg aH$); and second, that $H$ is nearly constant until nearly before inflation ends. Both premises are compatible and correct. However, the tensor modes might interact nearly the entire duration of inflation, therefore, any relevant scale must leave the horizon just about inflation ends, i.e., at $k^{(tensor)}\simeq H_{end}a_{end}$. Thus, our main methodology is not longer valid at this regime, so we must abandon this prescription and develop an improved one in order to present a much better explanation of any probable observational indication of this sector. Additionally, WI projects a superior description, so we must investigate this further within this scheme.

We expect this novel mechanism of SWs production brings new cosmological sources, for which no astrophysical source has been identified, and which may be searched for in future detectors. 

%
%
\appendix

\section{Tensor power spectrum}\label{appendix_a}
There are three polarisation modes: $+\,,\times\,,\textgoth{b}$; each one has a tensor power spectrum, and given that these are uncorrelated, the power spectrum of the GWs for all modes must be computed by:
\begin{eqnarray}
P_{Y} &=& \sum_{A=+,\times,\textgoth{b}} \frac{2 k^{3}}{\pi^{2}}\left|^{(A)}Y_{k}\right|^{2} = \sum_{A=+,\times,\textgoth{b}} \frac{2 k^{3}}{\pi^{2}}\left[(^{(A)}Y_{k}^{\mathbb{R}})^{2} + (^{(A)}Y_{k}^{\mathbb{I}})^{2}\right] \, \nonumber \\
\hspace{2cm} &=&  \frac{2k^{3}}{\pi^{2}}\left[(^{(+)}Y_{k}^{\mathbb{R}})^{2} + (^{(+)}Y_{k}^{\mathbb{I}})^{2} + (^{(\times)}Y_{k}^{\mathbb{R}})^{2} + (^{(\times)}Y_{k}^{\mathbb{I}})^{2} + (^{(\textgoth{b})}Y_{k}^{\mathbb{R}})^{2} + (^{(\textgoth{b})}Y_{k}^{\mathbb{I}})^{2}\right] \, \nonumber \\
\hspace{2cm} &=& \frac{2k^{3}}{\pi^{2}}\left[3 (Y_{k}^{\mathbb{R}})^{2} + 3 (Y_{k}^{\mathbb{I}})^{2}\right] = \frac{6k^{3}}{\pi^{2}}\left[ (Y_{k}^{\mathbb{R}})^{2} + (Y_{k}^{\mathbb{I}})^{2}\right] \, \nonumber\\
\hspace{2cm} &=& \frac{6 k^{3}}{\pi^{2}}\left|Y_{k}\right|^{2} \,,
\end{eqnarray}
where $^{(+)}Y_{k}=^{(\times)}Y_{k}=^{(\textgoth{b})}Y_{k}=Y_{k}$ since these three polarisations are independent to each other. 
\section{Initial conditions}\label{appendix_b}
In this appendix we present a well described methodology of the proper procedure of the initial conditions of the scalar and tensor sectors. To solve the differential equations analytically, the main premise we take is that $\mathbf{6 b/M_{P}=m\,H/\Lambda}$ (here $m\epsilon\,\mathbb{R}$) since $H$ is nearly constant until nearly before inflation ends. Also we consider the regime $k\gg aH$, when the scales are well inside the horizon. And we utilise the conformal time or proper time $\tau$ or $dt=ad\tau$ (and since $\dot{H}\simeq 0$ we have $1/a=-H\tau$). 
\subsection{Scalar modes: $Z_{k}$}
\begin{figure}[htbp] 
\includegraphics[scale=1.0]{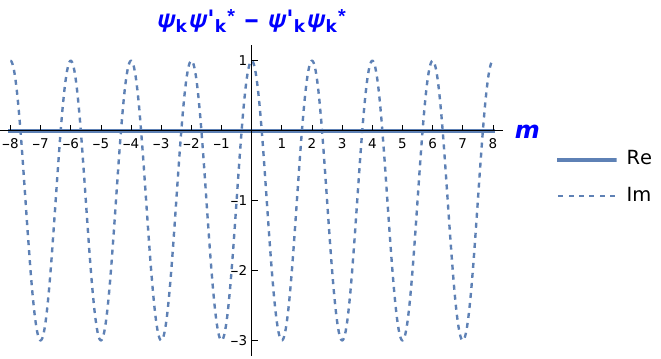}
\caption{Plot of the imaginary and real parts of $\psi_{k}\psi_{k}'^{*}-\psi_{k}'\psi_{k}^{*}$ with respect to $m$, here we take $\tau=-1$ and $k=1$, since numerically we normalise $aH\simeq 1$ so $\tau=-a^{-1}H^{-1}\simeq 1$. The real component is practically zero, while for even $m$ (positive or negative) 
the imaginary term is equal to $i$.}\label{fig:scalar_norm_condition}
\end{figure}

Consider eq.~(\ref{Zk}), which describes the scalar modes of the Fourier expansion of the field $\sigma$, having:
\begin{equation}\label{appendix_Zk} 
\ddot{Z}_{k}+(3-m)H\dot{Z}_{k} + \frac{k^{2}}{a^{2}}Z_{k} = 0 \,. 
\end{equation}
In order to solve above equation we introduce the variable $\psi_{k}=a Z_{k}$, therefore eq.~(\ref{appendix_Zk}) becomes:
\begin{equation}\label{appendix_psik} 
\ddot{\psi}_{k}+(1-m)H\dot{\psi}_{k} + \left(\frac{k^{2}}{a^{2}} + (m-2)H^{2}\right)\psi_{k} = 0 \,. 
\end{equation}
A more helpful form will be implementing the conformal time $\tau$, we also define $\psi_{k}'=d\psi_{k}/d\tau$ and $\psi_{k}''=d^{2}\psi_{k}/d\tau^{2}$. Therefore the evolution of Fourier modes $\psi_{k}$ in conformal time reduces to:
\begin{equation}
\tau^{2}\psi_{k}''+m\tau\psi_{k}' + \left(k^{2}\tau^{2}+(m-2)\right)\psi_{k} = 0 \,.   
\end{equation}
which solution is:
\begin{equation}\label{psik_full_solution}
\psi_{k}(\tau) = c_{1}t^{-(\nu+1)}\,J_{\nu}(k\,\tau)+c_{2} t^{-(\nu+1)}\,Y_{\nu}(k\,\tau) \,,  
\end{equation}
here $J_{\nu}$ and $Y_{\nu}$ are the first and second kind Bessel functions with parameter $\nu=(m-3)/2$. The constants $c_{1}$ and $c_{2}$ can be fixed by the normalisation condition: 
\begin{equation}
\psi_{k}\psi_{k}'^{*}-\psi_{k}'\psi_{k}^{*} = i \,,
\end{equation}
where by choosing:
\begin{equation}\label{c1_c2_values}
c_{1} = -\frac{\sqrt{\pi}}{2}k^{-m/2} \,,\quad c_{2} = i\frac{\sqrt{\pi}}{2}k^{-m/2} \,,
\end{equation}
we have 
\begin{equation}\label{psik_normalised}
\psi_{k}(\tau) = -\frac{i^{1-m}\sqrt{\pi}k^{-m/2}}{2}(-\tau)^{-(\nu+1)}\, H^{(2)}_{\nu}(k\, \tau)\,,
\end{equation}
here $H^{(2)}_{\nu}$ is the Hankel function of the second kind. Note that the domain of the normalised solution eq.~(\ref{psik_normalised}) is at $\tau<0$. The figure~\ref{fig:scalar_norm_condition} shows the evolution of the imaginary and real parts of $\psi_{k}\psi_{k}'^{*}-\psi_{k}'\psi_{k}^{*}$ with respect to $m$. The normalisation condition is satisfied for even $m$ (positive or negative). Moreover, table~\ref{tab.initial_conditions_functions} shows distinct expressions of $\psi_{k}(\tau)$ described by distinct values of $m$. Note that for $m=0$ the resulting $\psi_{k}(\tau)$ is quite similar to the Bunch-Davies vacuum mode function.
\begin{table}[h!]
\centering
\begin{tabular}{| c | c |}
\hline
$m$ & $\psi_{k}(\tau)$ \\
\hline
$2$ & $-\frac{e^{-ik\tau}}{\sqrt{2k} k\tau}$ \\
\hline
$0$ & $\frac{e^{-ik\tau}}{\sqrt{2k}}\left(\frac{1 + i k\tau}{k\tau}\right)$\\
\hline
$-2$ & $\frac{e^{-ik\tau}}{\sqrt{2k}}\left(\frac{-3 - 3ik\tau + k^{2}\tau^{2}}{k\tau}\right)$ \\
\hline
$-4$ & $\frac{e^{-ik\tau}}{\sqrt{2k}}\left(\frac{15+15ik\tau - 6k^{2}\tau^{2} - ik^{3}\tau^{3}}{k\tau}\right)$ \\
\hline
$-8$ & $\frac{e^{-ik\tau}}{\sqrt{2k}}\left(\frac{945+945ik\tau - 420k^{2}\tau^{2} - 105ik^{3}\tau^{3} + 15k^{4}\tau^{4} + ik^{5}\tau^{5}}{k\tau}\right)$ \\
\hline
\end{tabular}
\caption{Table of various expressions of $\psi_{k}(\tau)$ described by distinct values of $m$. Note that for $m=0$ the resulting $\psi_{k}(\tau)$ is quite similar to the Bunch-Davies vacuum mode function.}
\label{tab.initial_conditions_functions}
\end{table}

\subsection{Tensor modes}
\begin{figure}[htbp] 
\includegraphics[scale=1.0]{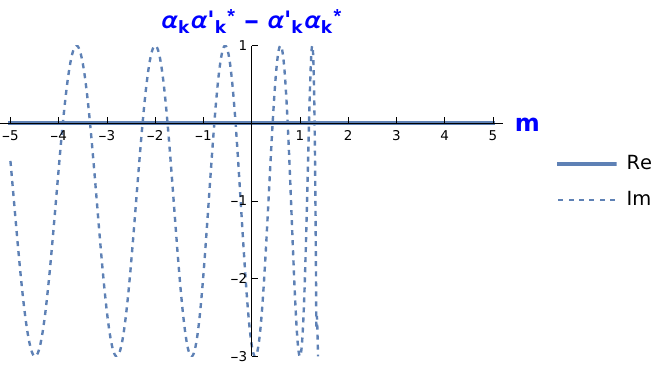}
\caption{Plot of the imaginary and real parts of $\alpha_{k}\alpha_{k}'^{*}-\alpha_{k}'\alpha_{k}^{*}$ with respect to $j$, here we take $\tau=-1$ and $k=1$, since numerically we normalise $aH\simeq 1$ so $\tau=-a^{-1}H^{-1}\simeq 1$. The real component is practically zero, while for instance with $m=-2$ the imaginary term is equal to $i$.}\label{fig:tensor_norm_condition}
\end{figure}

The tensor sector is treated in the same way as the scalar one. We consider eq.~(\ref{Y_k}) and once again we take $\mathbf{6b/M_{P}=m\,H/\Lambda}$ (as well as $\dot{H}\simeq 0$), having:
\begin{equation}\label{appendix_Yk} 
-\ddot{Y}_{k} + \left(1+m\right)H\dot{Y}_{k} + \left[4(1-m)H^{2} - \frac{k^{2}}{a^{2}}\right]Y_{k} = 0 \,, 
\end{equation}
To solve above equation we introduce the variable $\alpha_{k}=aY_{k}$, hence eq.~(\ref{appendix_Yk}) becomes:
\begin{equation}\label{appendix_Tpsik} 
-\ddot{\alpha}_{k}+(3+m)H\dot{\alpha}_{k} + \left[\left(2-5m\right)H^{2} - \frac{k^{2}}{a^{2}}\right]\alpha_{k} = 0 \,.
\end{equation}
Once again we use the conformal time or proper time $\tau$ or $dt=ad\tau$ ($1/a=-H\tau$), therefore the evolution of Fourier modes $\alpha_{k}$ in conformal time reduces to:
\begin{equation}
-\tau^{2}\alpha_{k}''-\left[4+m\right]\tau\alpha_{k}' + \left[2-5m-k^{2}\tau^{2}\right]\alpha_{k} = 0 \,,    
\end{equation}
which solution is:
\begin{equation}\label{psik_full_solution}
\alpha_{k}(\tau) = \tilde{c}_{1}\tau^{-(3+m)/2}\,J_{\mu}(k\,\tau)+\tilde{c}_{2} \tau^{-(3+m)/2}\,Y_{\mu}(k\,\tau) \,,  
\end{equation}
where $J_{\mu}$ and $Y_{\mu}$ are the first and second kind Bessel functions with parameter $\mu=\sqrt{m^{2}-14m+17}/2$. The constants $\tilde{c}_{1}$ and $\tilde{c}_{2}$ can be fixed by the normalisation condition: 
\begin{equation}
\alpha_{k}\alpha_{k}^{'\,*}-\alpha^{'}_{k}\alpha_{k}^{*} = i \,,
\end{equation}
where by choosing:
\begin{equation}\label{tilde_c1_c2_values}
\tilde{c}_{1} = -\frac{\sqrt{\pi}}{2}k^{m/2} \,,\quad \tilde{c}_{2} = i\frac{\sqrt{\pi}}{2}k^{m/2} \,,
\end{equation}
we have 
\begin{equation}\label{alphak_normalised}
\alpha_{k}(\tau) = -\frac{i^{1-m}k^{m/2}\sqrt{\pi}}{2}(-\tau)^{-(3+m)/2}\, H^{(2)}_{\mu}(k\, \tau)\,,
\end{equation}
here $H^{(2)}_{\mu}$ is the Hankel function of the second kind. Note that the domain of the normalised solution eq.~(\ref{psik_normalised}) is at $\tau<0$. The figure~\ref{fig:tensor_norm_condition} shows the evolution of the imaginary and real parts of $\alpha_{k}\alpha_{k}'^{*}-\alpha_{k}'\alpha_{k}^{*}$ with respect to $m$. Note that the normalisation condition is satisfied for instance for $m=-2$, yielding:
\begin{equation}
\alpha_{k}(\tau) = \frac{e^{-ik\tau}}{\sqrt{2k}}\left(\frac{-15i + 15k\tau + 6ik^{2}\tau^{2}-k^{3}\tau^{3}}{k^{4}\tau^{4}}\right) \,,\qquad m=-2 \,.  
\end{equation}

\acknowledgments
This work was supported by the CONAHCYT Network Project No. 376127 {\it Sombras, lentes y ondas gravitatorias generadas por objetos compactos astrofísicos}. R.H.J is supported by CONAHCYT Estancias Posdoctorales por M\'{e}xico, Modalidad 1: Estancia Posdoctoral Acad\'{e}mica and by SNI-CONAHCYT. C.M. wants to thank PROSNI-UDG support.

%
%
\bibliographystyle{unsrt}
\bibliography{gsbr}

\end{document}